\documentstyle[12pt]{article}
\begin{document}
\def\thebibliography#1{\section*{REFERENCES\markboth
 {REFERENCES}{REFERENCES}}\list
 {[\arabic{enumi}]}{\settowidth\labelwidth{[#1]}\leftmargin\labelwidth
 \advance\leftmargin\labelsep
\usecounter{enumi}}
 \def\newblock{\hskip .11em plus .33em minus -.07em}
 \sloppy
 \sfcode`\.=1000\relax}
\let\endthebibliography=\endlist

\hoffset = -1truecm
\voffset = -2truecm


\title{\bf  From Nuclear  To  Sub-Hadronic  Physics : A Global View  Of Indian Efforts  
}
\author{
{\normalsize\bf
Asoke N Mitra \thanks{e.mail: ganmitra@nde.vsnl.net.in}
}\\
\normalsize }
\date{}
\maketitle

\begin{abstract}

A panoramic view seeks to  trace  the principal ideas 
governing the phenomenal growth of Nuclear Physics from a 
modest beginning in the Thirties, to an  all-embracing field now
protruding to  the  subhadronic on the one  hand,  and to high 
temperature QGP  physics on the other.  In this narrative, 
which makes no claims to completeness, Indian efforts have 
been  highlighted  vis-a-vis   other similar efforts,   within  a 
common standard of global recognition.       

\end{abstract}

\section{Birth Of Nuclear And Particle Physics}

Although Nuclear Physics was born more than a century ago, 
with the discovery of Radioactivity ( A.H. Becquerel  1894) ,  its 
modern avatar dates back to   1932 ,  with  the
discovery of the {\bf Neutron} by J. Chadwick. It was then that the
$proton-neutron$ composition  of the $nucleus$ became known
for the first time. The next landmark event was the discovery by
Hideki Yukawa (1935)  of the short-range forces (the $meson$) that
hold the nuclear particles ($nucleons$) together by very
short-range forces, so as  to account  for Rutherford's observation on
the tiny size of the nucleus compared to its electronic jacket.
The relative size problem was aptly summarized by Hans Bethe's
pin-head analogy. Namely, the nuclear size bears roughly the same
ratio to the atomic size as does a pin-head to a standard sized
living room. Further, the atomic size itself bears roughly the
same ratio to the size of the pin-head ! However, Yukawa's
candidate for the short-range forces was not immediately
available, but only a "wrong" candidate  in Carl Anderson's  discovery 
(1937) of the $muon (\mu)$ which interacted weakly with matter, but
which was later to revolutionanize our concept of weak interactions
 (see below). The real breakthrough occurred  only a decade later, 
with the discovery by  Cecil Powell (1948) of the $\pi$-meson ($pion$) 
which showed  the right property of "Strong Interactions" as envisaged 
by Yukawa.  This formally marked the birth of $Particle$ Physics, but 
in retrospect, the dynamic role of the $pion$ as  the mediator of  
strong interactions  within  the framework of modern Quantum  Field 
Theory ($QFT$),  was to come in perspective much later (see below).

\section{ From Nuclei To Quarks : An  Overview }

Since the thirties, the initial growth of nuclear physics had been  rapid, in 
terms of both experimental and theoretical ideas. Experimentally, several
classes of accelerators (Cockcroft and Walton, Van de Graeff and
Cyclotron, as well as  their numerous updates) were  invented in quick succession as 
sources of energetic particles for the study of nuclear reactions
in the laboratory,  together with techniques for the detection of the
reaction products (counters, cloud chambers, etc).  The early work on 
nuclear reactions quickly established the short range character of nuclear
forces and the small size of the nucleus. On the other 
hand, the electron spectrum of  beta-decay was to open up still another 
front in unravelling the mysteries of short-range nuclear interactions,
this time at the level of $weak$ interactions ! First, it  led to Pauli's 
ansatz of the $neutrino( \nu_e)$  as a $new$  massless and electrically 
neutral particle, to account for the observed characteristics of the 
$\beta$-spectrum. This in turn led to  Fermi's formulation of the $correct$ 
theory of beta-decay, one in which the  electron and its associated neutrino
( $\nu_e$) were to play the role of a fundamental pair $(e, \nu_e)$ forming 
a  $weak$ interaction current ! A second pair $(\mu, \nu_\mu)$ was to come.
\par
In the meantime theoretical efforts had been under way since the
mid-Thirties (at the hands of Wigner, Bethe, Heisenberg, Pauli,
Peierls, and others) at organizing some principal features of
nuclear interactions, in terms of certain general principles of
Symmetry and identity among the nuclear constituents. Thus while Pauli's  
neutrino hypothesis paved the way to  an understanding of  $weak$ interactions, 
Heisenberg's contribution consisted in recognizing the key role of $exchange$
forces in the two-nucleon interaction, which is rather simply
realized by the $charged$ nature of the pionic force. Bartlett in
turn proposed another variant, viz., the $spin$-dependence of the
two-nucleon force. These individual features were subsumed in the
all-embracing idea of Wigner on the role of $Symmetry$ via Group
structures which provided a real breakthrough into the nature of
$strong$ interactions. This symmetry manifested as a rotational
invariance of the nuclear force within the space of $SU(4)$ which
encompasses two $SU(2)$ subspaces representing $spin$ and 
$isospin$ respectively. This important principle was to form the
bedrock of all subsequent microscopic theores  of nuclear forces.
\par
In  a major development two decades after the Fermi theory of beta-decay,   
the Lee-Yang theory of parity violation in beta-decay was quickly confirmed
by the experiment of C.S.Wu, eventually giving rise to a new theory of
weak interactions with emphasis on $handedness$, rather than parity, as 
the conserved quantity (cent per cent violation of parity !). In the jargon of 
Particle Physics, this  was called "$CP$" conservation, where "C" and "P"
stand for charge conjugation (particle-antiparticle interchange) and parity
(left-right) reversal, respectively. The  foundations of this new theory, 
although originally suggested by Hermann Weyl in 1929, were laid in a
fresh light  by Sudarshan--Marshak, Feynman--Gell-Mann, and Landau-Salam.
The Weyl theory which was thus  vindicated after initial rejection (on grounds
of "sanctity"  of parity conservation),  brought to the fore the traditional conflict 
of opposites  in  Physics, this time as a fight between Truth and Beauty [S. 
Chandrasekhar,  Physics Today 1969], when the latter eventually won, although 
not at the cost of the former.   And in this extended framework  of Physics, the 
(hitherto unwanted)  Anderson  $muon$,  together with its neutrino partner  
$\nu_\mu$,  found a natural place in  parallel with  the ($e, \nu_e$) pair, 
eventually leading to the modern theory  of weak  interactions now known 
as the  Glashow-Salam-Weinberg model. [For a  perspective  on the early phases  of the weak interaction
 sector,  see  P.K. Kabir,  "The CP Puzzle: Strange Decays of $K^0$" , Academic Press London  1968].

\subsection{ The Pion as the Trigger of Nucleon Resonances}

As  the harbinger  of strong interactions on the other hand,  the $pion$  turned
out to be the king-pin controlling the evolution of nuclear physics. While its
theoretical status in the nuclear domain  as a key element  in strong interaction  
physics has been  ever on the rise, its immediate  role as an $experimental$ 
probe into the nuclear interior  was almost instantly realized by the classic
experiments (in early 50's) of H.L. Anderson and his team at the U. of Chicago 
accelerator, shooting pion beams at hydrogen-like targets. The immediate fall-out 
was dramatic : the existence of short-lived peaks, termed $resonances$, soon 
to break up  into their original constituents, nucleons and pions, subsequently 
to be called  $hadrons$, together with their "strange" partners (see below).  
The first and most  prominent peak was the so-called "33" resonance, or 
$\Delta$, about 300 MeV above the nucleon,  to be followed by several others 
in fairly quick succession, so that the energy field above the nucleon was 
soon to be strewn with many nucleon-like but short-lived  peaks ! And as the
energy of the incoming pion beam increased, several more meson-like
states ($\rho$, $\omega$, $\phi$) also could be identified via more
elaborate analysis. Indeed in a rather short time span, (less than a decade
since the start of the Chicago experiments), the entire nucleon-meson
($hadron$) scenario was dotted with short-lived states spanning energies
up to a $GeV$, (i.e., almost the nucleon energy),  above the nucleon's. 
To add to the chaos, still another nucleon-like species ($\Lambda$)
and a corresponding pion-like meson ($Kaon$) were added to the
original nucleon-pion club, as a gift  from Cosmic Ray studies.  
And soon their corresponding higher lying short-lived states were to
be manufactured  by higher energy beams in the laboratory, leading
to an increase in Wigner's group classification from $SU(4)$  to $SU(6)$ "hadrons".    
\par
The question then arose: Should these short-lived peaks qualify as 
elementary particles on par  with nucleons and pions ? Considering the 
fact that they were found  to be highly unstable, breaking up into  their  
stable, lower energy species, they looked more like what atomic excited 
states were to their ground  state counterparts.  As if to strenghten  this view,  
the seminal electron scattering experiments on hydrogen targets by 
Hoffstadter et al (1957-) at the Stanford Linear Accelerator, showed clear 
structure effects usually associated with $composite$ bodies, in the 
hitherto believed `elementary' protons and neutrons whose experimental 
signatures showed up as their "charge" and "magnetic" form factors. 
  And now  their $strange$ ($\Lambda$ and $K$-like) counterparts  led to the
extended concepts of "strange baryons and mesons " under 
the generic  name of $hadrons$.  The situation now was strongly 
reminiscient of the pre-Bohr stage of the atom, this time at
the sub-nuclear level, suggesting a third generation of elementarity.

\subsection{ Third Generation of Elementarity: Quarks}

At this stage it  made  more sense to think of substructures of 
$hadrons$  than to give them elementary particle status ! It was at this 
time that the $quark-model$ of Gell-Mann (1964) and Zweig (1963)
was born, with curious properties of invisibility and fractional charges.
The quark-model, which has never looked back since then, has now
grown into a non-abelian  "gauge"  theory of the Yang-Mills type
with a multicomponent attribute called "Color" ,  under 
the name of  Quantum Chromodynamics, QCD for short, on similar
lines to  the single attribute (charge) gauge theory  known as QED, 
with quarks and gluons taking the place of electrons  and photons 
respectively. QCD however has properties complementary to QED,
namely, while the interaction strength of QED decreases with 
increasing distance, that of QCD  increases with increasing distance,
a property which goes under the name of {\it confinement} ! The opposite is 
true  for decreasing distance. In particular,  the QCD coupling strength 
$decreases$  with decreasing distance, a property that goes by the name
of $asymptotic freedom$. This unique property of QCD as  a non-abelian theory 
was first shown by Gross-Wilczek and Politzer (1973) and has been so well
confirmed by experiment, that these 3 authors were honoured by the
Nobel Prize for 2004 ! [ Fuller  details   are given  in the Article by 
Rajasekharan on the  developments  in Particle Physics].  
\par  
It was therefore inevitable that such fundamental substructures as 
quarks and gluons should penetrate deep into the domain of 
Nuclear Physics, and carve out a new sector, "hadron physics", 
as a subdiscipline of the latter. In this respect,  the demarcation 
between Nuclear and Particle Physics, which has never
been sharp, has been continually shifting ground, with the traditional 
nuclear range $(1-100 MeV)$ slowly extending to the few $GeV$ regime 
and even beyond, while the particle physics range is continuously  yielding place to the
former by shifting further and further to the right in tandem ! [This trend
is best reflected in the shifting PACS Classification Scheme over 
the years]. 

\section{ Growth Of Nuclear Physics: 3 Sectors }

The growth of NP against this huge background has necessarily been
multidimensional, since the subject touches several distinct sectors
from atomic to hadronic, involving  their mutual interactions from strong
to electroweak.  And of course each sector has generated its own
techniques (both theory and experiment), all within the general
head   that has come to characterize  NP. To get an idea of the
complexity of NP w.r.t. its atomic counterpart,  the biggest difference comes
from a  lack of quantitative knowledge of the precise nature of the  former,
vis-a-vis the latter (whose structure is better understood).  Indeed, the biggest
stumbling block on the  progress on the strong interaction front had
for long been the tentative status of the two-nucleon interaction throughout
the Fifties, leading M.L. Goldberger (considered an authority on
nuclear theory) to make a typically Churchillian remark: "in no
other field has so much effort been wasted towards so little
effect". As if to overcome this obstacle, Hans Bethe invoked the
$Second Principle$ theory according to which the $N-N$ interaction
was to be taken as the basic starting point for formulating the
applications to nuclear physics problems ( see further below). 
\par
It is such broad principles as above that have  characterized  the 
phenomenal growth  of NP  over the last 5 decades, necessarily in 
tandem with its related fields from atomic to sub-hadronic.  For  a
rough classification,  the principal  divisions  are : A) "classical" Nuclear 
Physics;
B) Pion-Nuclear Physics and  Quark Model; C) Nuclear  Astrophysics and 
QGP  transitions. 
Each  division (with further subdivisions)  has had a liberal sprinkling of
further  principles  or  "Models"  at  different levels of theory. The
scope of this  narrative is however  limited: the emphasis is  mainly on  those areas 
where  $Indian$  contributions ( working inside and outside the country) 
have  been  globally visible  during the last half century.   
      
\section{   "Classical"  Nuclear Physics : Shell Model }  

Nuclear Physics came of age after a  major  break-through  which occurred 
with the landmark discovery by Maria G Mayer  and Hans D. Jensen (1949) 
of the $Spin-orbit$ force in the nuclear  shell structure with an $opposite$ 
sign to that obtaining  in the  atomic shell structure. This  result, which had a 
firm basis in Wigner's  great  idea  of Symmetry in nuclear forces, gave a  
concrete  shape to the $nuclear shell model$, (despite the high density of a 
nucleus),  albeit  with an $inverted$  order w.r.t. the atomic 
shell structure,  and provided a natural understanding of several 
properties of nuclei,  far beyond the "magic numbers".  In particular, 
some general results emerged,  using  the symmetry between "particle"  and  
"hole" states in the shell model,  which was christianed as the
Pandya Theorem by J.B. French of Rochester, as outlined below.

\subsection{ Applications of the  Shell Model: Pandya Theorem }

The Nuclear Shell Model proved a strong impetus for more concrete 
formulations of Angular Momentum Theory (Elliot and Skyrme, 1951-52)
greatly extending Racah's pioneering work.  An important feature of such
formulations was that "particle" and "hole" configurations play
closely analogous roles. This is still another aspect of "Symmetry" 
at the nuclear level which closely resembles the more  familiar 
concept of particle-antiparticle symmetry ,  dating back
to Dirac's hole theory for the electron field. To see the analogy 
more closely,  let us recall that the `vacuum' in the electron's field 
corresponds to the state in which all negative energy states are filled.
And if enough energy (more than $2mc^2$) is pumped into this state,
one of the negative energy electrons jumps into a positive energy state,
creating a `hole' in the vacuum; then  the new state looks like  an
(excited) electron-positron (particle-hole) configuration. Analogously in a 
nucleus,  the `vacuum' state  is the `ground' state in which  all the 
`shells' are filled; this condition is just met in the  so-called 
"magic nuclei".  But if  enough energy is pumped into such a nucleus,
so that one of the nuclear particles (a nucleon) is raised into a higher
shell,  then one has an (excited) particle-hole configuration, just like
in the electron case ! And just as the electron-electron symmetry in
the former case has its counterpart in the electron-positron symmetry,
similarly the nucleon-nucleon symmetry in the nuclear case has its
counterpart in the  nucleon-hole  symmetry. As a word of caution, 
however, the analogy must not be carried too far : For while in the 
electron  field case, the `hole' (positron) is a genuine $antiparticle$  
(positron), the `hole' in the nucleon configuration is $not$ a true
antinucleon, and behaves like one only for the limited purpose
of `going in and out of  closed shells'.  [ the  energy scales
involved are in the  $MeV$ vs $GeV$ ranges respectively]. 
Even within this limited applicability, the particle-hole concept
in nuclear physics holds out strong correlative powers as
we shall see below.   
\par   
     From an applicational point of view, the
situation in the mid-fifties was that  shell-model calculations
for nuclear energy levels were possible only for a few 
$active$ nucleons in the configuration space of as many active
orbits. These were called the `major shells' of which the primary
ones of physical interest were $s-d$ and $f-p$. For such
calculations, the primary input was a two-particle matrix
element  of the form
$$ < j_1 j_2 J \mid H \mid j_1' j_2' J>,  $$
in the notation of a $jj$ configuration. Thus many calculations
were made at the beginning of a major shell ($s-d$ or $f-p$). In
this context, Gulio Racah had established a sort of $symmetry$
between $particles$ and $holes$ in a shell. This important
symmetry could thus be exploited to provide a connection between
two-particle and two-hole configurations so that results for the
former would be automatically applicable to the latter types.
\par
Now came the question of a possible connection between a
2-particle configuration and one with 1-particle plus 1-hole. This
important question was asked (and answered) by a young Indian
physicist Sudhir Pandya working with J.B. French at Rochester.
Using an extension of the symmetry argument of Racah, he showed
that indeed the matrix elements of 2-particle and 1-particle plus
1-hole configurations are related by an equation of the form
$$ < j_1 j_2^{-1}J\mid V \mid j_1 j_2^{-1}J> = \sum_x (2x+1)
W(j_1 j_2j_1 j_2 J x) <j_1 j_2 x\mid V \mid j_1 j_2 x> $$
connected by Racah coefficients. Pandya applied this result to 
the lowest four levels of $^{38}cl$
and $^{40}K$ where only a few spin-parity assignments had been
made. This gave rise to concrete predictions which were testable.
J.B. French (who called it the $Pandya Theorem$), got   the Utrecht 
nuclear physics group  to verify this result to within $~10 KeV$ ! 
This was a most
impressive confirmation, considering the fact that the usual
limits of agreement with experiment were taken as the order of
$100 KeV$, because of the uncertainties of nuclear matrix
elements which are in the $MeV$ regime. More significantly, 
this confirmation also suggested  that the states involved were 
almost pure $d_{3/2} f_{7/2}$ and $d_{3/2} f_{7/2}^{-1}$ states ! 
Further, by assuming the wave function to be "pure" (no configuration 
mixing), one could set an  upper bound to the contributions of 3-body 
forces to the energies of nuclear states ! Indeed the Pandya result 
was the first reliable and quantitative estimate of 3-body forces
within the framework of the shell model. It also provided a very
useful tool for extending shell model calculations across shells,
for systems involving $both$ particles and holes. Indeed when
giant resonances were discovered and described in terms of
collective particle-hole excitations, the value of this result was
quickly realized. 

\subsection{Collective Excitations, Pairing Correlations}

The Pandya theorem is a good illustration of the richness of
information forthcoming from a judicious use of subtle
symmetry principles connecting vastly different sectors of 
nuclear systems.   Another early discovery in  nuclear  structure  was  the large 
quadrupole deformation of  certain  classes of nuclei,  leading to the 
emergence of new, $collective$,  degrees of freedom( Bohr--Mottelson). 
These  were basically collective  excitations (rotations,  vibrations) of  
non-spherical nuclei, leading to a 
greatly simplified description of the energy levels of $deformed$  
nucle, albeit in a highly "macroscopic"  form.  To give a more 
$microscopic$ description of  these ideas,  was a big challenge that
was to extend over a few decades involving a synthesis of the collective
and nuclear shell models. This effort has had a long history, one
in which  the concept of $pairing$ (correlations 
between  nucleons  coupled to "zero spin") proved crucial,  as it  led to a simple 
understanding of the  energy gap between even-even and  odd-odd
nuclei, in close  analogy to the BCS Theory of Superconductivity in 
the  totally different  area  of  condensed matter physics.  Later,  the 
pairing  idea got a major boost in the form of the "Interacting Boson
Model (IBM)" (Arima--Iachello, 1974),  in which the d.o.f.'s
are those of effective zero spin $bosons$ made up of two nucleons 
of opposite spins. This boson-like character in turn proved mathematically
equivalent  to $paired$  fermion-like states, through a suitable mapping
process of both "unitary" and "non-unitary"  types.  These ideas have  
been   extensively   developed on a global scale, with a view to a 
more concrete synthesis of the (macroscopic) collective model with 
the (microscopic) single-particle shell model. In this respect, Indian
participation has been significant,  with  Y. K. Gambhir  having 
established a one-one correspondence between the Shell Model
states and the second quantized version of the basis states in the
Quasi-particle theory.  

\subsection{Nuclear Reactions : Direct Interaction Theory} 

In parallel with these breakthroughs  in nuclear structure, there had
been equally impressive progress  in the dual sector
of nuclear reactions,  whose understanding could be classified
in terms of i) a two-step process: resonance formation and  decay independent
of the formation mode; and ii) a one-step process: reaction time comparable 
to the passage of the projectile through the nuclear volume. 
The celebrated example of the first type is the Breit-Wigner 
one-level formula (subsequently generalized to more levels, like the
Kapur-Pieirls formula ), while 
the second type could be described by  an effective "optical potential" 
with  real and imaginary parts.  Seminal work in the second category was carried out by 
(the late) Manoj Banerjee and Carl Levinson who developed the Direct Interaction theory of inelastic scattering, using realistic shell-model states 
under $SU(3)$ classification,  
wherein  a `direct interaction'  is postulated to take place anywhere inside the nuclear volume, with 
remarkable agreement with data (on carbon) over a wide energy range.   
\par
The '70's and '80's   were a period of consolidation for a quantitative
 understanding of nuclear physics with
improvement in experimental techniques and a matching growth 
in computer technology to support the theoretical framework. 
New experimental techniques included the use of i) heavy ion beams
to investigate new features in nuclei ; ii) electron beams for mapping
precise nuclear-charge distributions (CEBAF) on the lines of
Hofstadter's original experiments ; iii) new forms of accelerators 
using  superconductivity; iv) intense pion and kaon beams 
under the name of "meson factories" to study specific features of
nuclear properties in precise details; and v) $K^-$ beams targetted 
at chosen nuclei, to produce a new class of nuclei,  termed 
"hypernuclei",  in which a long lived  baryon ($\Lambda$) is 
bound with the conventional nucleons.                     

\subsection{ Microscopic Models  And  Nuclear  Structure}

The Bethe Second Principle Theory  proved a strong incentive to the 
formulation of more microscopic models of nuclear structure in terms
of the two--nucleon  interaction taken as basic.  This is  best exemplified 
by the Bethe-Brueckner theory of infinite matter
using systematic approximations like the Hartree approximation with 
simplified forms of the basic nucleon-nucleon forces which
leads to the "mean field" in which nucleons move.  However even before the 
the full ramifications of the former,  some limited  but  important issues like the  
understanding of  spin-orbit forces in nuclei,  in terms of the well-known 2-body tensor 
force as input,  had engaged the attention of nuclear theorists. In particular,  the magnitude 
of this tensor force  was found inadequate for understanding the strength of the spin-orbit 
force governing the Shell-Model. This led Robert Marshak (with his student Signell) to 
propose a primary spin-orbit force (apart from  the usual tensor force) to account for 
its role in nuclear structure. B.P. Nigam and M.K. Sundaresan (1958-59) showed that 
this `direct' spin-orbit force could indeed account for the magnitude of the shell-model 
spin-orbit force.  Other  applications of a similar type  such as  the calculation of  nuclear  
energy  levels  in terms of simplified two-body forces (Mitra-Pandya 1960),  were generalized  
by Kramer-Moshinsky (1960) for the matrix elements of two-body forces in c.m. and  
relative coordinates,  later to be known as Moshinsky Brackets.  

\subsubsection{ Hugenholtz-VanHove (HVH) Theorem}

For more substantial applications of microscopic methods,  the  Bethe-Brueckner theory  
had to face the  rigours of  the Hugenholtz-Van Hove (HVH) Theorem  which  deals with the  
single particle properties  of an infinite interacting Fermi gas at absolute zero of temperature, and yields   
a relation amongst Fermi energy (E) , the average energy (A) per particle and the 
pressure of the system. However for a saturating system at  equilibrium, the
pressure vanishes, and one gets a simple relation $E=A$, i.e., for  an interacting Fermi 
system in the ground state, its average energy becomes equal to the Fermi energy ! 
This is a rare theorem in many-body physics which  is rigorously true up to all order of 
perturbation. It is also   remarkable that the theorem does not depend upon the precise 
nature of interaction,  and  is in general valid for any interacting self--bound infinite 
Fermi system, and  in particular for nuclear matter. It  was with  this theorem that  
Hugenholtz  and Van Hove found internal inconsistency in the early nuclear matter 
calculation of  Brueckner(Phys. Rev. 110(1958)597).  More dramatically,  the HVH
theorem did not spare even   the  celebrated   Bethe--Weiszacker Mass  formula 
which had been conceived  in a "classical"   manner,   as was first pointed out 
by Satpathy and Patra [Phys.Rev.Lett.51(1983) 1243], using a generalized HVH theorem 
for infinite nuclear matter  in which  a distinction was
made between the numbers  of  protons as well as their respective fermi energies.  

\subsubsection{Extreme Nuclear States}

The  work of Satpathy et al, [ Phy. Rep. 319(1999)85] 
 had important ramifications  on  several  new areas of NP ,   
especially those nuclei which are involved  near the limits of nuclear binding.  
According to the usual wisdom based on Wigner's symmetry
energy considerations, for a fixed number of protons not more
than a certain maximum number of neutrons can be bound,
and vice versa for a fixed number of neutrons. These "drip
lines"  are of interest in the laboratory, since nuclear properties 
may change drastically near these. More especially, they 
are of interest in (high $T$) stellar processes where a sequence 
of captures takes place rapidly.  Exploration of these limits
is a comparatively recent phenomenon (since 1990's). 
One interesting phenomenon  that occurs along the drip line
is "proton radioactivity" where the nuclei are "dripping" protons
since their binding energy is insufficient , but the Coulomb barrier 
retards their emission.  Neutron excess nuclei are more difficult 
to reach in the laboratory; since neutrons 
see no Coulomb barrier, their density distributions extend far
beyond `normal'  nuclear radii, indeed well beyond the 
corresponding proton distributions in neutron rich nuclei. This
asymmetry in turn will cause a change in shell structure due
to a substantial reduction in the spin-orbit term. For all such nuclei, 
the Satpathy et al work  suggest the existence of 
{\it new magic numbers}  and associated new islands of stability which have 
profound  implications  on  the broadening of the stability peninsula. 

\subsubsection{Yrast states}

Another kind of extreme nuclear states are the so-called Yrast states which are
formed as follows :  compound nuclear states at high excitation are first produced by
medium energy ($~200 MeV$) projectiles of heavy nuclei like $^{48}Ca$ 
incident on targets like $^{120}Sn$. These decay rapidly by emitting neutrons,
leaving the product nucleus in a fast rotating  state with large  angular 
momentum. Now in a rotating frame, the excitation energy is not too 
 high wrt the $lowest$ energy that the nucleus can have at this 
angular momentum -- the so-called $yrast line$-- but under the 
influence of $centrifugal$ forces, the lowest configuration in the 
nucleus can be quite different from those at low angular momentum.
Stated differently, the shell structure of nuclei becomes distorted 
under the centrifugal effect of rotation, and new levels  of relative stability 
may develop as  functions  of quadrupole deformation.  As a result, there
appear new classes of nuclear states with high deformation, in which several
nucleons have individual quantum numbers that differ from those of 
the normal ground state.  

\subsubsection{Bethe-Brueckner Theory in Practice}

Despite the formal constraint of the HVH theorem,  the Bethe-Brueckner theory
proved a strong incentive to researchers in formulating new techniques for
nuclear  reactions and structure calculations within its framework. (The late) Manoj
Banerjee  in particular,    with his student Binayok Dutta 
Roy,   was one of the first to apply Brueckner's reaction matrix to a 
finite nucleus calculation. Other  Indian 
contributions in this field  are due to  Nazakatullah (TIFR) and M.K. Pal (SINP).

 \subsubsection{Hartree-Fock techniques}

Nuclear many body calculations within the Bethe-Brueckner framework were 
developed along two  distinct tracks: Hartree-Fock and Thomas-Fermi. 
Significant Indian contributions  in the former were made by Y.K. Gambhir,   
leading  to deviations from the mean  field approximation.  
Relativistic effects were also studied by  Y.K. Gambhir and C.S. Warke on 
the basis of the J.D. Walecka formalism.  Experimental techniques for 
detection of these 
micro-states include electron scattering as a convenient tool for 
measuring the individual shell-model orbitals as well as the corrections 
due to correlation effects  arising from the short range  N-N interaction. 
The V. Pandharipande group (1997)   at the Univ of Illinois  showed  substantial deviations from a 
mean field approximation  due to  pair correlation effects in the high 
momentum  regime, in agreement with  observation [ Rev Mod Phys. {\bf 69}, 981 (1997)].        

\subsubsection{Thomas-Fermi method}

The Thomas-Fermi  approximation in nuclei was developed by R.K. Bhaduri and
C.K. Ross (Phys.Rev.Lett 27, 606 (1971)) in the form of a systematic expansion
of higher order terms, termed the "Extended Thomas-Fermi (ETF) method" by
the authors. The latter has been applied  extensively in shell corrections 
(in fission) and in the  construction of mass formulae.

\subsection{Few--Body Systems}

The  Bethe Second Principle Theory, which has  been central to  microscopic formulations 
of the nuclear many-body problem,  despite the constraints  of the HVH Theorem,
came in particularly useful for the (much smaller) "few-body"  systems whose  
mandatory quantum mechanical formulation would automatically satisfy such
constraints. Historically,  the  importance of  such systems may be traced back to
the  traditional insolubility of the atomic helium 3-body problem which stemmed from the non-linearity inherent in the Coulomb interaction, a 
disease which was not cured by quantum modifications, but had to wait for another 70 years for the concept of `periodic orbits' in classical non-linear systems to come to the rescue ! [See Introduction, Sect 3 for essential logic]. Of course,  the traditional  solution which has since held centre-stage in physics,  was the the famous variational principle  proposed by Hyllaraas  to tackle that problem. The
success of this principle therefore came in handy  for the $nuclear$ 3-body problem 
where the ground state energies of $^3 H$ and $^3 He$ were  quite fashionable topics
of research in the initial stages of NP.  The stumbling block was however the  use of
variational wave functions which would  hide the  micro-causal  structure of the 3-body wave 
function that cannot be obtained without solving  the corresponding 
Schroedinger equation . For the atomic helium problem there was no choice, since the basic (Coulombic) structure of   the two-electron force was not negotiable, but  other options were available   for the nuclear 2-body potential  which  was at best parametric,
thanks to the Bethe Second Principle Theory.  This option   could  now be
turned to  a  practical  advantage since  the  `effective range theory' (valid at low energy)
could fix only two constants--the scattering length and effective range, and many `potentials'
could be constructed with these two parameters in place. Y.  Yamaguchi (1954) took 
advantage of this choice by using `separable potentials' which would make the
Schroedinger equation exactly soluble, and produce an "exact" wave function without
variational uncertainties. Mitra [Nucl Phys 32, p529, 1962] and his colleagues (B.S. Bhakar 1963,  
V.S. Bhasin 1963, V.K. Gupta 1965)  took this result a step further by proposing that a separable
2-nucleon potential would reduce the (insoluble) 3-body problem to an ordinary 2-body
problem so that its microcausal structure would no  longer  be held hostage  to  variational 
uncertainties. In the initial days  of  computer  technology, this was a big help for a 
physical understanding  of the 3-body system via  its rich  structure, whose  information  
is  effectively   coded in the $off-shell$ matrix elements of the 2-nucleon potential.         
The idea caught on rather quickly on a global scale, as an intensely practical way to 
handle the 3-body  problem with its rich  applicational field,  at just the right time 
when Faddeev  proposed his  3-body formalism,  since  the  full-fledged  structure
of  a 3-body  wave function (with its crucial off-shell ramifications)  was free from the uncertainties of  variational wave  functions.  Gillespie-Mitra- Panchapakesan- Sugar (1964) soon extended the formalism to the 4-body problem, but the precise 
mathematical significance  of the Yamaguchi separable potential  became
clear: the $N$-body nuclear problem with separable potentials gets exactly reduced
to the level  of an ordinary  $(N-1)$-body problem !   With the rapid increase in computer technology,
the `separable' potential ansatz got  generalized to separable expansions for arbitrary
potentials, and the few-body problem soon got industrialized with bigger teams working 
on more realistic  2-body interactions involving both scattering and bound state problems.

\subsubsection{Three-Body Forces}

As an offshoot of the Bethe Second Principle Theory, the  $N-N$ interaction 
soon acquired some allied concepts since this effective formulation was not
microscopic enough  for  handling higher order effects, unlike, e.g.,   a  standard 
Yukawa-type interaction mediated by a field.  A popular concept in this regard
arose under the name of "Three-body forces",  wherein $three$ nucleons are
simultaneously involved in the interaction ! The concept can be formulated
in several ways ranging from a purely empirical contact ($\delta$-function)
interaction, to elaborate diagrammatic forms involving two pion exchanges 
(in succession) among the 3 participating nucleons; [the latter necessarily 
involved the folding of the Yukawa-type  pion- nucleon vertex].   Special care was
needed to ensure that such diagrams are truly "irreducible", i.e.,  they
are not part of the iterations of the input two-nucleon interactions. This in
turn causes uncertainties in the precision of  estimates  which often vary.
This field has generated much literature. prominent names being  Bruce 
McKellar and R. Rajaraman, but is still largely open-ended. 
     
\section{  Pion-Nuclear Physics \& Quark Model}

These areas represent the  main off-shoots  of "classical"  NP,  because
of the growing importance of new emerging degrees of freedom symbolized
by the $pion$ on the one hand, and the $quark$ on the other. 
Indeed   the $Quark Model$  of the $hadron$ (nucleon and meson) came as
an `answer to prayer' for the practitioners of traditional nuclear physics 
  since it was almost  tailor-made as a $few-body$ 
scenario for these nuclear  constituents at the $quark$ level !  It may be recalled that the
quark model had been evolving  in $two$ independent directions,
viz., a) as primary $currents$ obeying the laws of $SU(3)\times
SU(3)$ algebra; and b) as $constituents$ of hadrons in much the
same way as $hadrons$ are constituents of nuclei. [The $SU(3)$ is
a reminder of the inclusion of strange ($s$) quarks in the same
package as non-strange ($u,d$) $SU(2)$ quarks]. It is the
constituent quark scenario (b) that almost fitted into the
few-body scenario at the quark level, while the `current' quark 
picture  was more  suited to the conventional language of currents for particle
physics; [see  Rajasekharan for the latter ].  An intermediate approach
was to formulate NP with a special status to the pion as an effectively
`elementary' particle  interacting with nuclear constituents, while in the
full-fledged  constituent  quark model  all nuclear constituents 
are  treated as quark composites.      

\subsection{Chiral Properties of the Pion}
 
The $pion$   as the smallest quark composite   has found  its  special role as a $bridge$ between hadron 
and  quark-gluon physics. This role of the pion owes its origin
partly to certain "exact"  formulations which incorporate its
solitonic  properties (the Skyrmion) even before the birth of the
quark-model. Secondly the insolubility of QCD led to certain
alternative formulations incorporating at least some  important 
properties  of QCD, especially $chiral invariance$. This means
that the quark-gluon Lagrangian is invariant wrt chiral transformations
in the limit of zero $ud$ quark masses. By eliminating the quark 
and gluon d.o.f.'s from a path integral formulation of QCD in favour
of pion-like fields having the Skyrmionic properties, Gasser and
Leutwyler (1984) derived an effective Lagrangian in the form of a
momentum expansion which was chirally invariant. This "chiral
perturbation theory"  which was thus rooted in QCD, was enormously 
successful in reproducing certain known pre-QCD results (such as the 
$\sigma$ model, as well as predicting correctly many low energy 
hadronic properties. And  an extension of these ideas to include 
vector and axial vector mesons has proved equally fruitful. This
area has been one of the favourite hunting grounds of applicational  hadron physics, one in which Indian workers have played
leading roles in association with their global counterparts.  In particular,     
(the late)  Manoj Banerjee was  one of the pioneers in 
realizing the importance of chiral symmetry in hadronic structure, and 
doing concrete calculations in the chiral soliton framework.   Other
Indian  workers who have made significant contributions in this field,
together  with their many western counterparts are Rajat Bhaduri, 
 C.S. Warke, Y. K. Gambhir and V. Devanathan,  
among the more prominent ones. In particular,  in the related field of muon-capture in nuclei,  V. Devanathan and  R. Parthasarathy  have done significant work on   recoil polarization and  gamma-neutrino  angular correlations in muon capture ( 1980-85) which motivated  D. Measday's experiments at   TRIUMF on muon capture by silicon-28, 
leading to more reliable estimates of the pseudoscalar coupling constant than obtainable from PCAC  estimates. Subsequently, Parthasarathy made more detailed studies  of  neutral current induced neutrino excitations in  Carbon-12  nuclei  which formed the basis of  the KARMEN 
 experiment on  neutrino oscillations.

\subsection{Constituent Quarks,  Statistics And  All That}

We now come to the constituent quark model which has been
patterned closely after the corresponding formulations of
nuclear physics in terms of  nuclear constituents, albeit at
the level of one and two nucleons in practice, and that too 
 in terms of  Bethe's  Second Principle Theory, despite the
existence of QCD as a formal strong interaction theory !   

\subsubsection{Quark Statistics}

An important prerequisite of any formulation with constituent quarks
is the statistics they are supposed to obey. By simple arguments, 
the apriori choice would naturally be Fermi statistics, taking the
known number of d.o.f.s into account (spatial, spin and flavour).
On the other hand there were both dynamical and observational
constraints  with this simple choice. "Dynamical"  because the minimal
fermionic state compatible  with  the (symmetric) $56$ representation
would be  spatially $antisymmetric$,  but this would go contrary to 
standard dynamical expectations (Gursey-Kuo-Radicatti 1964) which 
would  prefer the lowest (ground) state to be spatially symmetric.  
Observationally,  a spatially antisymmetric ground state of 3 quarks
should show $nodes$ in the proton e.m. form factor at  rather low 
momenta ( Mitra-Majumdar 1966), something 
which is not observed. To overcome the dynamical problem of  a
symmetrical wave  function,  Greenberg (1964) suggested a new 
form of  statistics termed $parastatistics$ which would yield a
symmetrical ground state as required. But such a hypothesis would
go  against the canonical Bose-Fermi alternatives,  and was soon 
overshadowed by the now familiar d.o.f. termed $color$ a la
Gell-Mann,  in addition to space-spin-flavour,  so that Fermi statistics
could be logically accommodated  in such an  extended Hilbert space.
Thus  the proton form-factor anomaly  turned out to be  an $experimental$  fore-runner
of the color hypothesis which was to be subsumed in  a non-abelian
gauge theory  (QCD) that was taking shape,  with color as the right 
candidate for the needed attribute [ see Rajasekharan for details]. 

\subsection{ Hadron Spectroscopy With Quarks}

The $uds$-flavoured  quark structure of hadrons gave rise to a
straightforward extension of  the isospin group from $SU(2)$ to
$SU(3)$  which, together with the spin group $SU(2)$, led to an
$SU(6)$ extension of Wigner's $SU(4)$ group for nuclei. This was
a most pleasing development for the classification  of 2- and 3-quark
hadrons in their  ground states,  while their  excited configurations 
needed a further extension to $SU(6) \times O(3)$  to take care of
internal orbital excitations. On the basis of  this group,  Dalitz [1966]
made a  detailed classification  of possible  hadron resonances, and 
most of the observed resonances fitted nicely with this pattern,
except for a few odd ones which would presumably require more
subtle  dynamical considerations. Further checks were provided
by their decay patterns, mostly to pions and nucleons,  for which
the principal mechanism is via single quark transitions 
[Becchi-Morpurgo 1965], in close parallel to nuclear  e.m. transitions.
The Becchi-Morpurgo mechanism proved extremely successful for
explaining most two-body decays of hadron resonances, thus 
providing powerful checks on the  (Dalitz-like) $SU(6)\times O(3)$
assignments on their expected energy levels on the basis of
a simple harmonic  oscillator model (Greenberg-Resnikoff 1966; Karl-Obryk 1968], 
except  for  certain anomalies  which remained unexplained. One such 
anomaly was the observation of  considerably enhanced modes 
for heavy mesonic ( $\eta$) decays compared with their pionic modes.
This was explained by taking account of "Galilean invariance" of the
single quark pion interaction [ Mitra-Ross 1967], pending a full-fledged 
relativistic treatment  later  [Feynman et al 1971].           

\subsubsection{$N-N$ Interaction  via  Quark Exchanges}  

Among other  approaches to hadronic interactions in terms of
the quark model,  considerable attention  was devoted to  the derivation 
of   the $N-N$ interaction in terms of  the (microscopic) $qq$ and 
$q{\bar q}$ interactions,  through a  basically  "folding"  process. The 
input  form  of  the latter, in turn,  would often be coulombic in the 
high momentum limit,  but  also one that would  incorporate color 
confinement in the low energy  limit ( a la Richardson 1977].  Prominent 
Indian workers who have used  these techniques  systematically,  
include C.S. Warke, R. Shankar,  B. K. Jain, and Y.K. Gambhir.  
 But practical  obstacles in the way  of relativistic  formulations 
(an essential requirement for $light$ quarks !) and 
mathematically satisfactory  solutions of the relativistic two- and 
three-   quark problems  impeded further progress.  A good 
perspective on  such limited studies  is offered by  the Book of  
R.K. Bhaduri:  Models of the Nucleon (Addison-Wesley. 1988).

\subsection{ Covariant  Salpeter Eq : Markov-Yukawa Principle }
       
 In this respect,  a  major formulation was  initiated by Feynman 
et al  [1971], FKR for short,   in a novel approach to hadron
structure which insisted on inclusion of (low energy) spectroscopy
on par with high energy processes in any serious quark level study
of the strong interaction problem, with Lorentz and gauge
invariance as integral parts of the underlying formalism. This was
to prove a big incentive for more systematic studies of the
few-quark systems (even 3 decades after $QCD$ had been 
formally christioned  as a strong interaction candidate). This coveted
project was taken up in the Eighties by the  Delhi group 
( D. S. Kulshreshtha, A.N. MItra, I.Santhanam, N. N. Singh]
in  a step by step manner,  using the Bethe-Salpeter Equation (BSE) as the
basic  dynamics, but with the crucial recognition that  the time-like degree 
of freedom must $not$ be taken on par with the space-like d.o.f.s (this lesson 
was learnt from  the difficulties of the FKR approach precisely in this
regard).  This led to a reappraisal of the Salpeter Equation (1952), based on the Instant Approximation to the BSE,  thus 
giving it a 3D content.  But the (remaining)  time-like information had to be  recovered 
by $reconstructing$  the  4D BS wave function in terms of 3D ingredients (This information was already there in the original Salpeter equation, 
but had apparently got lost in the subsequent literature), so as
to give a two-tier  structure : the (reduced) 3D   BSE form  attuned  to  hadron  
spectroscopy,  and its   (reconstructed) 4D form  ideal for  the calculation  of
various  hadronic transition processes  via Feynman diagrams  for 
appropriate quark loops.  On these lines an  extensive study  of hadron
spectroscopy  for  $q{\bar q}$ and  $qqq$  systems,   provided a unified  view of both  
spectroscopy  and transition  processes within a single unified framework.  
Relativistic  covariance was ensured by  giving  a 
special status to the c.m. frame of the concerned hadron (as quark  composite).
This   2-tier  Salpeter-like  framework, in turn,   found a  firm  theoretical 
basis in  a 50-year old  principle,  traceable to  Markov (1940)  and   Yukawa (1950),  
which came  to  be  known as the Markov-Yukawa Transversality Principle ($MYTP$).  
It   not only provideed  a Lorentz-covariant meaning to   this 
framework  for  a 3D-4D interlinkage of all
Salpeter-like equations via $Covariant Instantaneity$, but also  gave   it  a
more secure foundation through a  new  $gauge principle$ in which the
quark constituents interact  in a direction $transverse$  to the  momentum $P_\mu$
of the composite hadron.    
This was   already quite satisfactory for  hadron spectroscopy (which needed  a 3D
equation  only),  but  unfortunately gave rise to a `Lorentz  mismatch  disease ' for  
Feynman  diagrams  involving 4D  quark-loops,  (since  the different vertices
involved  different c.m.  frames),  leading to $complexities$ in the 
resultant  hadronic amplitudes !  To  cure this disease  needed a further 
strategy,  involving  Dirac's  light-front  formalism,  whose main ideas 
are outlined as follows :

\subsection{Dirac's Light-Front Dynamics  and MYTP}        

 Apart from laying the foundations of QFT through his famous
Equation, Dirac made two great contributions within a year's gap
from each other: a) light-front dynamics (LFD)[1949]; b) constrained
dynamics [1950].  The former  which is  the subject  of this 
discussion,  is characterized by  the following idea:  A  relativistically 
invariant Hamiltonian can be based on different classes of initial surfaces:
instant form ($x_0=const$); light-front (LF) form ($x_0+x_3=0$);
hyperboloid form ($x^2+a^2 <0$). The structure of the theory is
strongly dependent on these 3 surface forms. In particular,   the
"LF form" stays invariant under $7$ generators of the Poincare'
group, while the other two are invariant only under $6$ of them.
Thus the LF form has the maximum number ($7$) of "kinematical"
generators (their representations are independent of the dynamics
of the system), leaving only 3 "hamiltonians" for the dynamics.
\par
    Dirac's  LF dynamics got a boost after Weinberg's discovery of the
$P_z =\inf$ frame which greatly simplified the structure of
current algebra. The Bjorken scaling in deep inelastic scattering,
supported by Feynman's parton picture, brought out the equivalence
of LF dynamics with the $P_z=\inf$ frame.  The
time ordering in LF-dynamics is in the variable $\tau=x_0+x_3$, instead
of $t=x_0$ in the instant form.  Indeed,  the LF dynamics often turns out to be simpler and
more transparent than the instant form, while retaining the  net
physical content more efficiently.  The LF language was
developed systematically within the QFT framework  through
world-wide activity since the Seventies (Kogut, Susskind--1974).  Indian 
participation was visible  lately  (Srivastava (1999], Mitra (1999)] .

\subsubsection{ Covariant  Light Front Dynamics}

The final link  in the  FKR  programme  was achieved by endowing
the  $MYTP$  framework   with  a  "covariant light-front"   approach which
yields loop integrals free from time-like momentum anomalies. This was
checked from a recent application of $MYTP$ in  the  LF formalism to the Pion
Form Factor [Mitra, Phys Lett. B 1999]. The extension to the $qqq$ problem also  
goes through,  but is more technical [ Mitra-Sodermark hep-ph/0104219].

\subsubsection{ Heavy Quarkonia and Exotic States}

We end this section   with  some comments on   the  developments  since  the so-called
"November revolution" (1974) which led to  the  discovery of  Charmonium $c{\bar c}$,  as 
the first  example of  "heavy quarkonia",   to be followed (1977) by the discovery of a still heavier 
one ( `Beauty' $b{\bar b}$),  culminating in  the  identification of the `Top' at the end of the 
last Century.   This led to extensive activities  in heavy quarkonia spectroscopy  on both 
theoretical and experimental  fronts.  In these activities,  Indian  participation has  also been 
(expectedly) widespread.  Allied things  like  QCD exotic  states like $q{\bar q}$  hybrids 
and $gg$ glueballs, also received due attention.   Such states are characterized  by  more 
than the minimum number of quarks compatible with their basic quantum numbers,  yet  permitted by 
the quark model in principle)  which are hard to detect (!)    A prominent Indian group led by Kamal Seth  
participated in the  heavy quarkonia programmes in Fermilab as well as the CLEO
Collaboration for $e^+ e^-$  annihilation at Cornell.  Among his successes  were the detection 
of  $two$  rare  radially excited   pseudoscalar charmonium states.    

\subsubsection{ Hidden Color}

As a  closely  related item,  "hidden colour"  is a concept which arises from the colour 
confinement ideas in QCD and the quark model for the multi-quark
configurations like 6-q, 9-q, 12-q etc.  By the very nature of  such  configurations
(say 3n),  they have a direct bearing on the  corresponding nuclear configurations 
with numbers $n$, and it is of interest to ask if  certain  anomalies  pertaining
to  the  latter  can  be explained by this novel concept.  In this respect, some
notable Indian efforts  deserve particular mention.  Afsar Abbas  at IOP Bhubaneshwar
made a systematic  group theoretic study of the  amount of  hidden colour in 9-q and 12-q 
configurations ( Phys Lett 167B ((1986) 150),  and  used this  concept to explain the 
experimentally obserevd  "hole" in the matter distribution in the nuclei: 4-He, 3-He and 3-H.
Abbas  has   also found that the same concept is useful in explaining such diverse
properties as that of the "neutron halo" , nuclear clusters and nuclear  molecules 
( Mod Phys Lett A16 (2001) 735).  If  confirmed  by further  probes,  it  should 
warrant the interpretation  that  all these properties  represent  the signatures of the 
quark degrees of freedom in the ground state properties  of nuclei.   It is understood
that  experiments at KEK Japan and  GANIL France are under way for  testing these predictions. 

\section{ Nuclear  Astrophysics:  QGP  Transition }

An important area of NP  dates back to its astrophysical applications 
 which were  pioneered by Hans Bethe (1939) through his analysis of $p-p$ chain 
and carbon-nitrogen  cycles for turning hydrogen into helium.  Although in a 
dormant state for some  decades, it has come to life  once again for more than 
one reason. The study  of neutrinos from the Sun  has acquired a new dimension 
when the traditional mismatch of expected rates compared to the 
Davis  result,  has started getting  resolved only recently after 
the latest Kamiokande results were put in perspective with
the expectations from  "neutrino mixing" !  A more important  
reason for the present tendency of nuclear  physics to 
interface with  the cosmological domain has been triggered by
the use of heavy  ion beam reactions,  so as  to facilitate the 
transition to the  "Quark-Gluon-Plasma" (QGP) state, and thus    
help create early universe conditions  of high temperatures 
in the laboratory.  Although this programme is stll at an early
stage, it has already brought together two hitherto disparate
groups (nuclear and plasma physicists) towards a common 
goal  of re-enacting a phase transition-- that between the 
hadronic and the QGP phases--in the laboratory. 
                 
\subsection{Matter at High $\rho$ and $T$: QGP Transition ?}

Closely related to the domain of Nuclear Astrophysics,  is  matter 
at high densities ($\rho$) and  temperatures ($T$) which naturally 
exists in the stars, but the task of  simulating  such conditions 
in the laboratory represents a  major challenge which has 
become a subject of intense interest  during the last 2 decades, 
involving both theoretical and experimental activities in tandem.  
Now theoretical calculations based
on temperature-dependent field theory (QCD) for high densities
(incorporated through  suitable chemical potentials) suggest
that  under conditions of high $T,\rho$ in a volume large
compared to a typical hadronic volume,  a transition occurs 
to a new state of matter where quarks are no longer confined to 
their individual hadrons, but can move freely within the
larger volume. This state which is called the quark-gluon-plasma
(QGP) state,  illustrates the intertwining of two hitherto distinct
branches of physics-- nuclear and plasma physics-- brought about
under conditions of high temperature. For illustration, we  present  
a brief sketch of   a sustained  investigation by 
Subol Dasgupta (McGill) and his  colleague George Bertsch  
in this regard.  

\subsection{Boltzmann-Uehling-Uhlenbeck Model}

Dasgupta and Bertsch  developed a transport model for heavy  ion 
collisions at  intermediate energies,  termed the BUU (Boltzmann-
Uehling-Uhlenbeck) model, which has been in use  for the last 
two decades,  and has  the following essential elements.  In a  collision 
of  one nucleus against another, in the lowest approximation, 
each nucleus can be regarded as a collection of ``frozen'' nucleons. 
Taking first  a hard scattering model  for simplicity,  when
the two nuclei pass through each other, one nucleon from a nucleus
will scatter off another nucleon from the other nucleus provided
the impact parameter between the two nucleons
is less than $\sqrt{\sigma(s)/\pi}$, $\sigma(s)$  being the cross section for
a 2-nucleon collision at c.m. energy $s$. If the nuclei are  moderately large, after
a few collisions the nucleons will  not remember  which nucleus
they came from !  One then  finds the experimentally testable result that after a time of the
order of   $10^{-24} sec$, the ordered motion has changed, the
collisions have ceased and nucleons are streaming freely. 
\par 
In an alternative scenario,  a nucleus by itself stays bound  in a
 mean field ( a  la  Hartree-Fock theory).  Two such nuclei  can  be boosted 
towards each other. When they are passing through each other, any
nucleon will feel the mean field of all other nucleons, not just the one
they originally belonged to (  S. Koonin). An  advantage
of this scenario is  that it  is not  necessary  to consider  the nucleons 
as " frozen"  before collisions (unlike the BUU model), and 
Fermi motion is easy   to  incorporate. 
\par
DB  were able to combine   both the scenarios by  replacing  Hartree-Fock
by Vlasov propagation,   thus facilitating  the  use of Vlasov equations  which are 
familiar in  electron plasma theory.  The resulting  model has been used to 
i) put narrow limits on the  compressibilty of nuclear matter from heavy ion data; 
and ii)  predict   production of particles like pions and photons.
[G. F. Bertsch and S. Das Gupta, Physics Reports 160(1988)189]

\subsection{ Other Indian contributions}

Among the more indigenous Indian contributions in this area,  those by
B. Banerjee,  J. Parikh, L. Satpathy, C.P. Singh  and B. Sinha  are among 
the more prominent ones.  Parikh's  publications (including 3 books) reflect 
his simultaneous  expertise in the  dual fields of nuclear and plasma physics, 
wherein  he has  proposed 
a signature for testing formation of QCD plasma:  look for 
pion correlations in relativistic heavy ion collisions. [This was  a year earlier  than the
famous Matsui-- Satz  proposal on the J/PSI suppression].
\par
Satpathy has successfully employed a diatomic-like molecular model 
(termed Dynamic Potential Model)  in describing the
phenomena of nuclear molecular resonances observed in heavy-ion
collisions. 
\par
Singh, whose group is a member of the experimental PHENIX Collaboration 
at BNL,    has made extensive studies of the signals of QGP
 in terms of his proposed equation of state,  and found some unique 
signatures in the $strangeness$ sector, which are being actively 
looked at by experimentalists probing QGP. 
\par  
Sinha's notable  contribution in  relativistic heavy ion collisions  lies  in 
the recognition  of single photons and dileptons as  signatures of QGP.
which are being looked for  at CERN.  

\subsection{Experiments at RHIC:  Results from 4  Groups}

To explore this state,  experiments  have been carried out 
successively at Bevalac (Berkeley), AGS (Brookhaven),
SPS (CERN), and now at the  new Relativistic Heavy Ion 
Collider (RHIC) at Brookhaven. Since the QGP  phase
transition is associated with quark $deconfinement$,
the central issue is how to observe and interpret  this 
feature unambiguously.  So far, in the earlier colliders
before RHIC, no clean signatures have been found, but  the latest outcome from RHIC (which has an order of magnitude
higher energy), is quite positive. This is indicated by the results from four groups analysing the data from two beams of gold ions clashing at several interaction zones around the ring shaped facility : 1) BRAHMS; 2) PHENIX; 3) PHOBOS; 
AND 4) STAR , all in Nucl Phys. {\bf A 757}, (2005).    Their consensus is that the fireball is a liquid of strongly interacting quarks and gluons, rather than a gas of weakly interacting ones. The liquid is dense, but seems to flow with very little viscosity, approximating an ideal fluid ! The RHIC  findings were reported at the April 2005 APS meeting in  Tampa. 
    
\section{ Retrospect : More Maths For  Nucl Physics }

The foregoing is merely a bird's eye view of the phenomenal growth of Nuclear
physics during the last 75 years,  whose scope has shrunk further  by  the  
emphasis on Indian contributions  where  relevant in the Global context.  This
growth has necessarily involved  a  dovetailing of Nuclear physics with other
major branches of physics, from particle physics to   quark model
QCD  and all the way to  early universe cosmology involving QGP
transitions,  bringing together their characteristic techniques. Thus  the 
investigations of the QGP  transition  has involved  an  active interaction of Nuclear
physics techniques with those of  Plasma physics (Vlasov eqs).   Similarly, 
the merging of  traditional Nuclear physics items with those of Particle physics 
has brought to the fore the need for $relativistic$ formulations with all their
mathematical sophistications such as  4D covariance and    
a new "gauge principle"   viz.,  the Markov-Yukawa Transversality Principle as a  
possible dynamical  scenarios for a joint solution of the 2- and 3-quark problems. 
\par
Finally a few words (not covered in the text) on the extended scope of nuclear studies to include highly 
mathematical quantum  techniques, especially the role of
newer  and newer methods of angular momentum  as well as the techniques 
of  Quantum Groups, for the  solution of diverse problems in  Nuclear  physics.  In this endeavour, Indian contributions have been quite prominent, 
especially those emanating from the MATSCIENCE  group of V. Devanathan, 
R. Jagannathan and K. Srinivasa Rao,  who have investigated such problems
in a most comprehensive fashion, with an impressive list of publications. 
\par
In conclusion,  it is fair to say  significant Indian contributions in this area 
(as in most others !) have been of a theoretical nature,  while experimental NP 
on Indian soil is still to get off the ground. On the other hand, Indian workers
have  generally shown a high degree of competence in working with bigger
Western groups, a necessary  condition for success in experimental NP. 
\par
I am grateful to  several Indian colleagues in India and abroad, for readily 
responding to my request for bringing their work to my attention. Any omissions
in this regard are entirely mine.  I have also benefitted  from the article
of  Henley and Schiffer in "More Things in Heaven and Earth" 
[ B.Bederson Ed, APS 1999].              

\end{document}